\documentclass[aps,showpacs,twocolumn]{revtex4}
\usepackage{graphicx}
\usepackage{dcolumn}
\usepackage{bm}
\input epsf
\usepackage{latexsym}
\begin{document}
\newcommand{\A}{{\mathcal{A}}}
\newcommand{\Od}{{\cal O}}
\newcommand{\dA}{\delta{\mathcal{A}}}
\newcommand{\oR}{\Omega_r h^2}
\newcommand{\og}{\Omega_{\gamma} h^2}
\newcommand{\om}{\Omega_m h^2}
\newcommand{\ok}{\Omega_k h^2}
\newcommand{\ob}{\Omega_b h^2}
\newcommand{\ode}{\Omega_{de} h^2}
\newcommand{\h}{\hat{H}}

{\bf Comment on "Viable singularity-free f(R) gravity without
 a cosmological constant"}

\vspace{0.5cm}

A modified $f(R)$ gravity model 
 has been recently proposed in \cite{quartin}
whose cosmological behaviour is clearly distinguishable from $\Lambda$CDM. 
Contrary to previous opinions which consider that self-consistent 
$f(R)$ gravity models distinct from $\Lambda$CDM are almost ruled out, the 
authors claim that the proposed model is cosmologically viable.
Here we show that although the model satisfies some consistency 
conditions, precisely because of its departure from $\Lambda$CDM behaviour,
it does not satisfy  local gravity constraints and, in addition,
the predicted matter power spectrum conflicts with 
SDSS data. 

Out of the four viability conditions  imposed on $f(R)$ theories
\cite{Silvestri_09_2007}, 
the proposed model satifies three of them. 
The fourth condition, namely,  
$\vert f_{R}-1\vert\ll 1$  at recent epochs,  is imposed 
by local gravity tests. Although it is still 
not clear what is the actual 
limit on this parameter, certain estimates \cite{HS}
give $\vert f_{R}-1 \vert <10^{-6}$ today, \cite{TS}.
 This condition also ensures 
that the cosmological 
evolution at late times resembles 
that of $\Lambda$CDM. However, in the proposed model,   
$\vert f_R -1\vert\sim 0.2$ today for $\alpha=2$ and 
$q_0\sim -0.25$. In principle, if we are only interested
in large scales,  we could ignore  
local gravity inconsistencies, but still the deviations from $\Lambda$CDM can have  
drastic cosmological consequences on the evolution
of density perturbations, as discussed by several authors \cite{Starobinsky,Bean,Cruz}. 

Thus, the linear evolution of  matter density perturbations for 
sub-Hubble ($k\eta \gg 1$) modes in $\Lambda$CDM is given by
the well-known expression:
\begin{equation}
\delta''+\mathcal{H}\delta' - 4\pi\text{G} \rho_0 a^2\delta\,=\,0
\label{delta_0_SubHubble}
\end{equation}
where $\delta=\delta \rho/\rho_0$, ${\cal H}=a'/a$ and prime denotes derivative 
with respect to 
conformal time $\eta$. Notice that the evolution of the Fourier modes
does not depend on $k$. This means that once the density contrast 
 starts growing after matter-radiation equality, the mode evolution 
only changes the overall normalization of the matter power-spectrum $P(k)$, 
but not its shape. 
\begin{figure}[h]
\begin{center}
{\epsfxsize=7.7cm\epsfbox{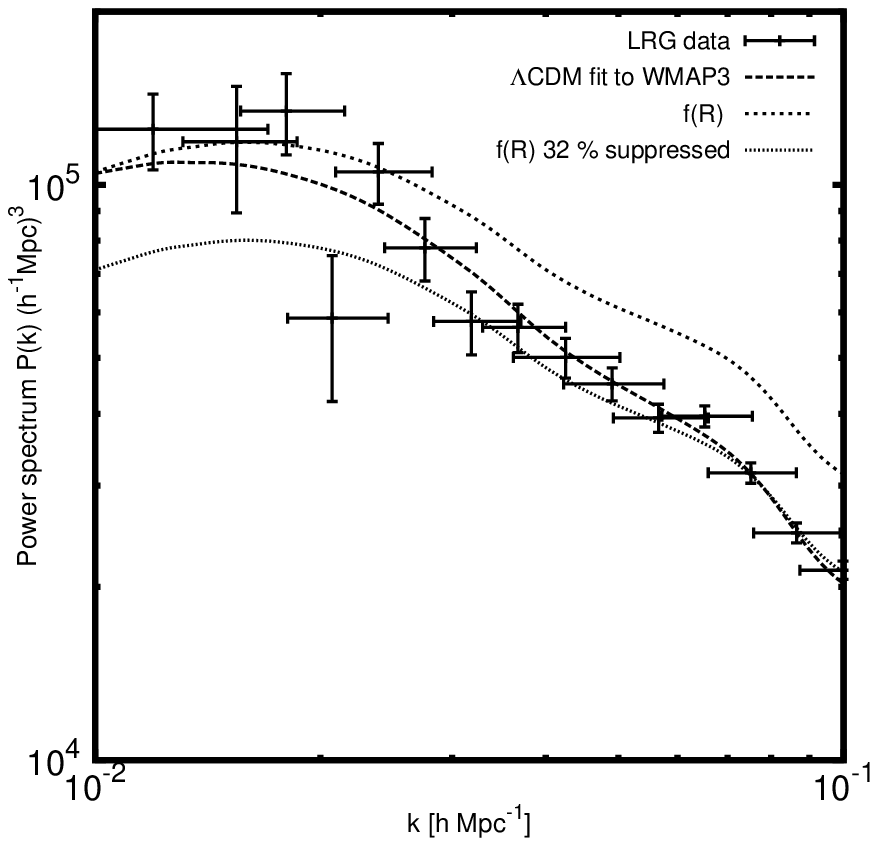}} \end{center}
{\footnotesize {\bf Figure 1:} Linear matter power-spectra for $\Lambda$CDM and  
$f(R)$ in \cite{quartin} with $\alpha=2$. Data from SDSS \cite{Tegmark}.}
\end{figure}
However in $f(R)$ theories, the corresponding equation reads \cite{Cruz}:
\begin{widetext}
\begin{eqnarray}
\delta''+\mathcal{H}\delta'+\frac{f_{R}^{5} \mathcal{H}^{2} 
(-1+\kappa_1)(2\kappa_1-\kappa_2)-\frac{16}{a^8}
 f_{RR}^{4}(\kappa_2-2)k^{8}8\pi \text{G} \rho_{0}a^2}
{f_{R}^{5}(-1+\kappa_1)+\frac{24}{a^8}f_{RR}^{4}f_{R}
(\kappa_2-2)k^{8}}\delta \,=\,0
\label{eqn_ours}
\end{eqnarray}
\end{widetext} 
where $\kappa_1={\cal H}'/{\cal H}^2$ and
$\kappa_2={\cal H}''/{\cal H}^3$. Notice the  $k^8$ dependence in the last
term which appears due to the fact that $f_{RR}\neq 0$. 
This means that the matter power-spectrum
is further processed after equality and the transfer function is modified
 with respect to that of $\Lambda$CDM.  
This drastically changes the shape 
of $P(k)$, as shown in Fig. 1, where normalization to WMAP3 has been imposed.
In the figure,  SDSS data from luminous red 
galaxies \cite{Tegmark}
and the  $\Lambda$CDM power spectrum from the linear perturbation theory with
 WMAP3  cosmological data 
are also shown. 
Notice that $\Lambda$CDM gives an excellent fit to data with $\chi^2=11.2$,
 whereas for the $f(R)$ theory  
$\chi^2=178.9$, i.e. $13\sigma$ out. Even if we drastically reduced 
the overall normalization 
in a 20$\%$, the 
discrepancy would still remain at the 7$\sigma$ level.  
Actually, the best fit would require a $32\%$ normalization reduction and 
still would be $4.8\sigma$ away (see Fig. 1).


A. de la Cruz Dombriz, A. Dobado and A.L. Maroto. Dept.
F\'{i}sica Te\'orica I, Universidad Complutense de Madrid, 28040
Madrid, Spain.


\vspace*{-0.6cm}

\begin{thebibliography}{0}
\vspace*{-0.7cm}
\bibitem{quartin} V.~Miranda, S.~E.~Joras, I.~Waga and M.~Quartin,
  Phys.\ Rev.\ Lett.\  {\bf 102}, 221101 (2009)
\bibitem{Silvestri_09_2007} L.~Pogosian, A.~Silvestri, Phys.\ Rev.\  
{\bf D77} 023503, (2008)
\bibitem{HS} W.~Hu and I.~Sawicki, Phys.\ Rev.\  D {\bf 76}, 064004 (2007)
\bibitem{TS} Difficulties with
chameleon-like mechanism have also been recently discussed in 
I. Thongkool, M. Sami, R. Gannouji and S. Jhingan, arXiv:0906.2460v1 [hep-th]
\bibitem{Starobinsky} A.A. Starobinsky, JETP Lett. {\bf 86}, 157 (2007) 
\bibitem{Bean} R.~Bean, D.~Bernat, L.~Pogosian, 
A.~Silvestri and M.~Trodden. Phys.\ Rev.\   {\bf D75} 064020, (2007).
\bibitem{Cruz} A.~de la Cruz-Dombriz, A.~Dobado and A.~L.~Maroto,
  Phys.\ Rev.\  D {\bf 77}, 123515 (2008)
\bibitem{Tegmark} M.~Tegmark {\it et al.} ,
  Phys.\ Rev.\  D {\bf 74}, 123507 (2006)

\end{thebibliography}
\end{document}